\documentstyle[12pt]{article}
\begin{document}
\setlength{\topmargin}{0.0cm}

\hfill LYCEN 9538

\hfill December 1995

\vspace{3.0cm}

\begin{center}
{\huge {\bf SOME ASPECTS OF}} \\
\vspace{0.5cm}
{\huge {\bf TRANSVERSITY$^*$}} \\
\vspace{1.5cm}
{\large X. Artru} \\
\medskip
{\it Institut de Physique Nucl\'eaire,
IN2P3-CNRS et Universit\'e Claude Bernard Lyon-I,
F-69622 Villeurbanne cedex, France  }
 \\

\vspace{2.4cm}

\large {\bf ABSTRACT}

\end{center}
The specificities of transverse polarization with respect to helicity of
ultrarelativistic fermions are pointed out.
For massless fermions, a covariant transversity four-vector is defined,
up to a kind of gauge transformation.
The tranversity distribution of quarks in a nucleon is defined.
Its possible connection to the magnetic or electric dipole moment of the
baryon is conjectured.
Consequences of the approximate chiral invariance
on transverse spin asymmetries in hard processes are enumerated.
The "sheared jet effect" introduced by Collins
for measuring the transverse polarization of a final quark is presented.

\section{What is transversity ?}

For a {\it massive} fermion, there is no problem of defining a
transversely
polarized particle of momentum $\vec p$ : put it at rest, polarize
it in some
direction $\hat n$ orthogonal to $\vec p$ and then apply the necessary
Lorentz boost to give it momentum $\vec p$.
For a {\it massless} fermion, this definition does not work
because there is no rest frame.
On the other hand, we know that {\it helicity} states exist and form
a complete
basis. A transversely polarized state should therefore be a linear
superposition of helicity ones.
To get the coefficients, the most natural method is
to give the particle a temporary small mass and let this mass go to zero.
Thus, for $\vec p$ along the positive $z$ axis, we get

\vspace{1.2cm}

\hrule

\smallskip
\noindent
$^*$contribution to the ELFE Summerschool and Workshop,
Cambridge, UK, July 22-28, 1995

\vfill

\pagebreak

\setlength{\topmargin}{-1.5cm}
\setlength{\textheight}{25.0cm}

\begin{equation}
|\hat n> = {|+> +\, e^{i\varphi} \ |->\over \sqrt{2}}
\end{equation}
where $\varphi$ is the azimuth of $\hat n$.

This receipe would not completely solve the problem in the case of a
chiral
symmetric world~: the relative phase of
the $|+>$ and $|->$ states would be arbitrary, which would make the
azimuth of
$\hat n$ ambiguous (a similar ambiguity would exist for linearly polarized
photons if the electric-magnetic duality were an exact symmetry).
Therefore the observability of transversity is linked to chiral symmetry
breaking.

In the massive case, any polarization of the fermion can be represented
in a noncovariant way by the 3-dimensional vector $\vec P=2 <\vec S>$
measured in the rest frame (i.e., before boost).
Boosting $\vec P$ results in the covariant polarization four-vector
\begin{equation}
{\cal S} = (0,\,\vec P_\perp) + P_L\, {1\over m}\, \tilde p
= {\cal S}_\perp + {\cal S}_L
\end{equation}
where $\tilde p=(|\vec p\,|, \, p^0\, \hat p)$ and
$\hat p\equiv \vec p/|\vec p\,|$.

When $m\to 0$, ${\cal S}_L$ becomes infinite
(unless $P_L$ is strictly zero).
However, the covariant projector $u(\vec p, s)\,\bar u(\vec p, s)$
keeps finite~\cite{Landau} :
\begin{equation}
u(\vec p, s)\,\bar u(\vec p, s) =
{1+\gamma^5\,\gamma\cdot {\cal S} \over 2}\,(\gamma\cdot p + m)
\to {1+\gamma^5\,\gamma\cdot {\cal S}_\perp + P_L\, \gamma^5
 \over 2}\,\gamma\cdot p
\end{equation}
This equation shows that in the massless case helicity and transversity
play
very different roles. The later is associated with one more $\gamma^\mu$
matrix
and is therefore called a {\it chirality odd}
quantity~\cite{Jaffe}.
Another interesting feature is the invariance under the
"gauge" transformation of the transversity four-vector
\begin{equation}
{\cal S}^\mu_\perp \Rightarrow {\cal S}^\mu_\perp
+ {\rm constant}\times p^\mu \,;
\end{equation}
this invariance makes the application of Lorentz transformations to
${\cal S}^\mu_\perp$ possible
(it would not be the case if we had imposed ${\cal S}^0_\perp \equiv 0$).
The "gauge" ${\cal S}^\mu_\perp = (0, \vec P_\perp)$ is the analogue of
the radiation gauge of the photon.
One may view the gauge freedom as the relic of an infinitesimal uncertainty
of the longitudinal polarization in the limit $m\to 0$.

\section{Transversity distribution inside the nucleon}

In analogy to the quark helicity distribution
$\Delta_L q(x) \equiv 2 g_1(x) = q^+(x) - q^-(x)$
in a longitudinally polarized nucleon $N^+$,
we define the quark transversity distribution~\cite{{Soper},{XAMM},{Jaffe}}
in a transversely polarized nucleon $N^{\uparrow}$ :
\begin{equation}
\Delta_\perp q(x) \equiv 2 h_1(x) = q^{\uparrow}(x) + q^{\downarrow}(x)
\end{equation}
It obeys the Soffer's inequality~\cite{Soffer}
\begin{equation}
|\Delta_\perp q(x)| \le q^+(x)
\end{equation}
which is stronger than the trivial one
$|\Delta_\perp q(x)| \le q(x)$.

$h_1(x)$ is not the same quantity as $g_1(x) + g_2(x)$,
in spite of the fact that the later is measured with transversely polarized
target. Only in a nonrelativistic quark
model do they coincide (but such a model is unrealistic for deep inelastic
reactions).
In the infinite momentum frame,
\begin{equation}
\Delta_\perp q(x) = {1\over 4\pi} \int dz <N^{\uparrow}|
\ \Psi_q^\dagger(0)\ (-\gamma^5\, \vec\gamma\cdot\vec P_\perp)
\ \Psi_q(0,0,0,z)\ |N^{\uparrow}> \ e^{-ik_z z}
\end{equation}
where $k_z=xp_z$ and $|N^{\uparrow}>$ is a nucleon plane wave
[normalized to $2p^0\,(2\pi)^3\,\delta(\vec p-\vec p\,')\,$]
with transverse polarization $\vec P_\perp$.
This quantity obeys the sum rule~\cite{{Jaffe},{Ralston}}
\begin{equation}
\int_0^1 \Delta_\perp \left[\,q(x) - \bar q(x)\,\right]\,dx =
\int\int\int d^3\vec X
<N^{\uparrow}|
\ \bar\Psi_q(\vec X)\ \vec\Sigma\cdot\vec P_\perp
\ \Psi_q(\vec X)\ |N^{\uparrow}> \,,
\end{equation}
where $|N^{\uparrow}>$ is now at rest and normalized to unity.
The right-hand side of (8) is called the {\it tensor charge}.
In fact, $\Sigma^i\equiv \epsilon_{ijk}\sigma^{jk}$ is part of the tensor
$\sigma^{\mu\nu}$). This tensor occurs in the anomalous magnetic interaction
${1\over2}\mu_a\ \bar\Psi\ \sigma^{\mu\nu}\ \Psi\ F_{\mu\nu}$,
therefore we may conjecture that the anomalous magnetic moment of the quark
contributes to the anomalous magnetic moment of the nucleon by
(tensor charge) $\times \mu_a$(quark). Similarly for electric dipole moments.
So, if we could tune these moments, one would get
\begin{equation}
{\partial \mu({\rm nucleon}) \over \partial \mu({\rm quark})}
= {\partial\, {\rm e.d.m.(nucleon}) \over \partial\, {\rm e.d.m.(quark})}
= \hbox{tensor charge}
\end{equation}

\section{Consequences of approximate chiral symmetry}

Transverse spin asymmetries are interferences between different helicity
amplitudes, the helicity difference being $\pm 1$. Two other equivalent
statements are

\noindent
- transverse spin asymmetries are helicity flip amplitudes of the
unitarity diagram
({\it e.g.}, $\Delta_\perp q(x)$ is the helicity flip amplitude
of forward $\bar q\,N\to\bar q\,N$ scattering).

\noindent
- transverse spin asymmetries correspond to
particle-antiparticle states of helicity $\pm 1$ in the appropriate t-channels
of the unitarity diagram ($N\,\bar N\to q\,\bar q$ for $\Delta_\perp q(x)$).

In deep inelastic reactions, quark masses can be neglected (if we discard
charm and bottom quarks) and the axial or vector interactions
with gauge bosons conserve helicity along the quark lines (chiral symmetry).
Combined with the previous statement,
this symmetry has the following consequences at leading twist
(i.e. to zeroth order in $m_q/Q$ or $m_h/Q$ )~\cite{XAMM}

- $a$) single spin asymmetries vanish

- $b$) there is no transversity correlation between external fermions not
belonging to the the same fermion line in the unitarity diagram :
"transversity information is confined in quark lines".

- $c$) the polarized cross section is invariant if we rotate
the $\vec P_\perp$ 's of all the external fermions simultaneously
by the same angle about their respective momenta
(we may restrict the rotation
to a subset of external fermions linked by fermion lines
in the unitarity diagram).

- $d$) gluons do not contribute to the transverse spin of the nucleon

- $e$) the transverse spin correlation between target and projectile vanishes
after integration over the azimuth of the final state.

Property $a)$ is significantly violated for instance in
$p+p\to\Lambda^{\uparrow}+X$ or $p+p^{\uparrow}\to\pi+X$,
but at medium $p_T$ ($\sim 1-2$ GeV) ;
the problem may become acute if it persists at larger $p_T$.
Property $b)$ forbids measuring $\Delta_\perp q(x)$ using polarized
beam and target in fully inclusive Deep Inelastic Lepton Scattering (DILS),
contrary to $\Delta_L q(x)$.
However, it allows $\Delta_\perp q(x)$ to be measured
in doubly polarized Drell-Yan experiment~\cite{Soper}
or in the semi-inclusive DILS~\cite{{Baldrachini},{Dourdan}}
\begin{equation}
e^-\ N^{\uparrow} \to e^-\ \Lambda^{\uparrow} + X\,;
\end{equation}
in the later case, one measures the polarization of the $\Lambda$ which is
proportional to
$\Delta_\perp q(x) \times \Delta_\perp D_{\Lambda^\uparrow/q}(z)$.
The second factor is the analysing power of the "quark polarimeter"
$q^\uparrow \to \Lambda^\uparrow + X$.

Property $c)$ gives for instance the
$1+A\cos[2\varphi_\mu-\varphi(\vec P_\perp)-\varphi(\vec P_\perp')]$
dependence of doubly polarized Drell-Yan.
It also imply property $e)$ as well as the relations
$A_{SS}=-A_{NN}$, $D_{SS}=D_{NN}$ between the double-spin asymmetry parameters.

\section{The sheared jet effect}

The fragmentation of a transversely polarized quark is {\it a priori}
not invariant by rotation about the quark momentum~\cite{Collins} :
\begin{equation}
D_{h/q^\uparrow}(z,\vec p_\perp)= D_{h/q}(z,|\vec p_\perp|) \times
\left\{1 + A_{h/q}(z,|\vec p_\perp|)\ |\vec P_\perp|
\ \sin\left[\varphi(\vec P_\perp) - \varphi(\vec p_\perp) \right]
\right\}
\end{equation}
This effect might be a more efficient quark transversity polarimeter than
$q^\uparrow \to \Lambda^\uparrow + X$. Both have to be calibrated at $e^+\,e^-$
colliders~\cite{{Chen},{calibrat}}.
The single spin asymmetry observed in
$p+p^{\uparrow}\to\pi+X$ may be a manifestation of it~\cite{trukiki}.

\section{Conclusion}
This very short introduction does not pretend to give an exhaustive view of the
present developments of the physics of partonic transverse spin.
It rather tries to bring out some particularities of this physics
compared to that of partonic helicity, and to convince the reader that quite
new information about hadron structure and chiral symmetry breaking
may be obtained from its experimental study.

\end{document}